\RequirePackage{ifpdf}
\ifpdf 
\documentclass[pdftex]{sigma}
\else
\documentclass{sigma}
\fi

\newenvironment{mybox}{\begin{array}{|c|}\hline} {\\\hline\end{array}}
\newcommand{\partiald}[2] {\cfrac{\partial #1}{\partial #2}\,}
\newcommand{\totald}[2]{\cfrac{\textrm{d}#1}{\textrm{d}#2}\, }
\renewcommand{\div}{\operatorname{div}}
\newcommand{\Div}{\operatorname{Div}}

\newcommand{\tr}{\operatorname{tr}}
\newcommand{\tp}{{\,\hspace{-2pt}^{\scriptscriptstyle T}}}

\newcommand{\adj}{\operatorname{Adj}}
\newcommand{\mysum}[2]{\displaystyle{\sum_{#1}^{#2}}}
\newcommand{\myint}[4]{\displaystyle{\int_{#1}^{#2}#3\;\textrm{d}#4}}

\newcommand{\point}{{\scriptscriptstyle\bullet}}

\newcommand{\R}{\mathbb{R}}

\newcommand{\vphi}{\varphi}
\newcommand{\x}{\boldsymbol{x}}
\newcommand{\y}{\boldsymbol{y}}
\newcommand{\z}{\boldsymbol{z}}
\renewcommand{\u}{\boldsymbol{u}}
\renewcommand{\r}{\boldsymbol{r}}
\newcommand{\s}{\boldsymbol{s}}
\newcommand{\w}{\boldsymbol{w}}
\renewcommand{\v}{\boldsymbol{v}}
\newcommand{\F}{\boldsymbol{F}}
\newcommand{\ggamma}{\boldsymbol{\gamma}}

\newcommand{\ovu}{\overline{\u}}
\newcommand{\ovp}{\overline{p}}
\newcommand{\ov}{\overline}
\newcommand{\ovs}{\overline{\mathsfb{S}}}
\newcommand{\ovw}{\overline{\mathsfb{W}}}
\newcommand{\ovdelta}{\overline{\delta}}
\newcommand{\tld}{\widetilde}

\DeclareMathAlphabet\mathsfb {OT1}{cmss}{bx}{n}
\newcommand{\tsr}{\mathsfb}   
\newcommand{\vt}{\boldsymbol} 
\newcommand{\scal}{\cdot}

\newcommand{\ttau}{\mathsfb{T}}
\renewcommand{\hat}{\widehat}
\newcommand{\e}{\mathrm{e}}
\newcommand{\RR}{\tsr{R}}
\newcommand{\J}{\tsr{J}}

\newcommand{\no}{{\itshape{\bfseries N}}}
\newcommand{\ye}{{\itshape{\bfseries Y$^*$}}}

\begin{document}
\allowdisplaybreaks

\renewcommand{\PaperNumber}{052}

\FirstPageHeading

\ShortArticleName{Consequences of Symmetries on the Analysis and
Construction of Turbulence Models}

\ArticleName{Consequences of Symmetries on the Analysis\\ and
Construction of Turbulence Models}

\Author{Dina RAZAFINDRALANDY and Aziz HAMDOUNI}

\AuthorNameForHeading{D. Razf\/indralandy and A. Hamdouni}

\Address{LEPTAB, Avenue Michel Cr\'epeau, 17042 La Rochelle Cedex 01, France} 
\Email{\href{mailto:drazafin@univ-lr.fr}{drazafin@univ-lr.fr}, \href{mailto:ahamdoun@univ-lr.fr}{ahamdoun@univ-lr.fr}}

\ArticleDates{Received October 28, 2005, in f\/inal form May 02,
2006; Published online May 12, 2006}

\Abstract{Since they represent fundamental physical properties in
turbulence (conservation laws, wall laws, Kolmogorov energy
spectrum, \dots), symmetries are used to analyse common turbulence
models. A class of symmetry preserving turbulence models is
proposed. This class is ref\/ined such that the models respect the
second law of thermodynamics. Finally, an example of model
belonging to the class is numerically tested.}


\Classification{22E70; 35A30; 35Q30; 76D05; 76F65; 76M60}

\section{Introduction}
Turbulence is one of the most interesting research f\/ields in
mechanics. But with the current performance of computers, a direct
simulation of a turbulent f\/low remains dif\/f\/icult, even
impossible in many cases, due to the high computational cost that
it requires. To reduce this computational cost,
 use of turbulence models is necessary. At the present time, many turbulence models exist
(see \cite{sagaut04}). However,  derivation of a very large
majority of them does not take into account the symmetry group of
the basic equations, the Navier--Stokes equations.

In turbulence, symmetries play a fundamental role in the
description of the physics of the f\/low. They reflect existence
of conservation laws, via Noether's theorem. Notice that, even if
Navier--Stokes equations are not directly derived from a
Lagrangian, Noether's theorem can be applied and conservation laws
can be deduced. Indeed, there exists a Lagrangian (which will be
called ``bi-Lagrangian'' here) from which Navier--Stokes
equations, associated to their ``adjoint'' equations, can be
derived. A way in which this bi-Lagrangian can be calculated is
described by Atherton and Homsy in \cite{atherton75}. The
expression of this bi-Lagrangian and the inf\/initesimal
generators of the associated Euler--Lagrange equations are given
in Appendix~A. However, the conservation laws are not studied in
this paper.

The importance of symmetries in turbulence is not limited to the
derivation of conservation laws. \"Unal also used a symmetry
approach to show that Navier--Stokes equations may have solutions
which have the Kolmogorov form of the energy
spectrum~\cite{unal94}. Next, symmetries enabled Oberlack to
derive some scaling laws for the velocity and the two point
correlations~\cite{oberlack01}. Some of these scaling laws was
reused by Lindgren {\it et al} in \cite{lindgren04} and are proved
to be in good agreement with experimental data. Next, symmetries
allowed Fushchych and Popowych to obtain analytical solutions of
Navier--Stokes equations~\cite{fushchych94a}. The study of
self-similar solutions gives also an information on the behaviour
of the f\/low at a large time~\cite{cannone05}. Lastly, we mention
that use of discretisation schemes which are compatible with the
symmetries of an equation reduces the numerical
errors~\cite{olver01,kim04}.

Introduction of a turbulence model in Navier--Stokes equations may
destroy  symmetry pro\-perties of the equations. In this case,
 physical properties (conservation laws, scaling laws, spectral properties,
large-time behaviour, \dots) may be lost. In order to represent
the f\/low correctly, turbulence models should then preserve the
symmetries of Navier--Stokes equations. The f\/irst aim of this
paper is to show that most of the commonly used subgrid turbulence
models do not have this property. The second goal is to present a
new way of deriving models which are compatible with the symmetries of
Navier--Stokes equations and which, unlike many existing models,
conform to the second law of thermodynamics. As it will be shown
in appendix,  conformity with this law leads to  stability of the
model, in the sense of~$L^2$.

The paper will be structured as follows. In Section 2, the
principle of turbulence modelling, using the large-eddy simulation
approach, will be concisely presented, as well as some common
models. These models will be analysed in Section~3 under the
symmetry consideration. In Section~4, a class of symmetry
preserving and thermodynamically consistent models are derived.
One example of a model of the class is numerically tested in
Section~5. Some conclusions will be drawn in Section~6. In
Appendix~A, it will be shown that Navier--Stokes equations can be
derived from a bi-Lagrangian. At last, in Appendix B, stability of
thermodynamically consistent models is proved.

\section{Large-eddy simulation}

Consider a three-dimensional incompressible Newtonian f\/luid,
with density $\rho$ and kinematic viscosity $\nu$. The motion of this
f\/luid is governed by Navier--Stokes equations:
\begin{gather}
\partiald{\u}{t}+\div(\u\otimes \u)+\cfrac{1}{\rho}\, \nabla p =\div \ttau,\nonumber\\
\div \u=0, \label{navierstokes}
\end{gather}
where $\u=(u_i)_{i=1,2,3}$ and $p$ are respectively  velocity and
pressure f\/ields and $t$ the time variable. $\ttau$ is a tensor
such that $\rho\ttau$ is the viscous constraint tensor. $\ttau$
can be linked to the strain rate tensor
 $ {\tsr{S}=(\nabla \u+\tp\nabla \u)/2}$ according to the relation:
\[
\ttau=\partiald{\psi}{\tsr{S}},
\]
$\psi$ being a positive and convex ``potential'' def\/ined by:
\[
\psi=\nu\tr \tsr{S}^2.
\]

Since a direct numerical simulation of a realistic f\/luid f\/low
requires
 very significant computational cost, (\ref{navierstokes})
is not directly resolved. To circumvent the problem, some methods
exist. The most promising one is the large-eddy simulation. It
consists in representing only the large scales of the f\/low.
Small scales are dropped from the simulation; however, their
ef\/fects on the large scales are taken into account. This enables
to take a much coarser grid.

Mathematically, dropping  small scales means applying a low-pass
f\/ilter. The large or resolved scales $\ov{\phi}$ of a quantity
$\phi$ are def\/ined by the convolution:
\[
\ov{\phi}=G_{\ovdelta}*\phi,
\]
where $G_{\ovdelta}$ is the f\/ilter kernel with a width
$\ovdelta$, and the small scales $\phi'$ are def\/ined by
\[ \phi'=\phi-\ov{\phi}. \]
It is required that the integral of $G_{\ovdelta}$ over $\R^3$ is
equal to 1, such that a constant remains unchanged when the
f\/ilter is applied.

In practice, $(\ovu,\ovp)$ is directly used as an approximation of
$(\u,p)$. To obtain $(\ovu,\ovp)$, the f\/ilter is applied to
(\ref{navierstokes}). If the f\/ilter is assumed to commute with
the derivative operators (that is not always the case in a bounded
domain), this leads to:
\begin{gather}
\partiald{\ovu}{t}+\div(\ovu\otimes \ovu) +\cfrac{1}{\rho}\nabla \ovp = \div (\ov{\ttau}+\ttau_s),\nonumber\\
\div \ov{u}=0, \label{navierstokesfiltrees}
\end{gather}
where $\ttau_s$ is the subgrid stress tensor def\/ined by
$\ttau_s=\ovu\otimes \ovu-\ov{\u\otimes \u}$ which must be
model\-led (expressed by a function of the resolved quantities) to
close the equations. Currently, an important number of models
exists. Some of the most common ones will be reminded here. They
will be classif\/ied in four categories: turbulent viscosity,
gradient-type, similarity-type and Lund--Novikov-type models.

\subsection{Turbulent viscosity models}

Turbulent viscosity models are models which can be written in the
following form:
\begin{gather}
\ttau_s^d=\nu_s\ovs, \label{turbulent_viscosity}
\end{gather}
where $\nu_s$ is the turbulent viscosity. The superscript ($^d$)
represents the deviatoric part of a tensor:
\[
\tsr{Q}\ \mapsto\
\tsr{Q}^d=\tsr{Q}-\cfrac{1}{3}(\tr\tsr{Q})\tsr{I_d},
\]
where $\tsr{I_d}$ is the identity operator. The deviatoric part
has been introduced in order to have the equality of the traces in
(\ref{turbulent_viscosity}). In what follows, some examples of
turbulent viscosity models are presented.
\begin{itemize}\itemsep=0pt

\item {\it Smagorinsky model} (see \cite{sagaut04}) is one of the
most widely used models. It uses the local equilibrium hypothesis
for the calculation of the turbulent viscosity. It has the
following expression:
\begin{gather*}
    \ttau_s^d=(C_S\ov{\delta})^2\vert \ovs\vert\ovs,
\end{gather*}
where $C_S\simeq 0.148$ is the Smagorinsky constant, $\ov{\delta}$
the f\/ilter width and $\vert
\ovs\vert=\sqrt{2\tr\big(\ovs^2\big)}$.

\item In order to reduce the modelling error of Smagorinsky model,
Lilly~\cite{lilly92} proposes a dynamic evaluation of the constant
$C_S$ by a least-square approach. This leads to the so-called {\it
dynamic model} def\/ined by:
\begin{gather}
    \ttau^d_s=C_d\ov{\delta}^2\vert \ovs\vert\ovs, \qquad \text{with}\quad C_d=\cfrac{\tr (\tsr{LM})}{\tr \tsr{M}^2}.
\label{dynamic}
\end{gather}
In these terms,
\[
\tsr{L}=\tld{\ovu}\otimes\tld{\ovu} -\tld{\ovu\otimes\ovu}, \qquad
 \tsr{M}=\ov{\delta}^2\tld{\vert\ovs\vert \ovs} - \tld{\ov{\delta}}^2 \vert \tld{\ovs}\vert \tld{\ovs},
  \]
the tilde represents a test f\/ilter whose width is
$\tld{\ov{\delta}}$, with $\tld{\ov{\delta}}>\ovdelta$.

The last turbulent viscosity model which will be considered is the
structure function model.

\item Metais and Lesieur \cite{metais92} make the hypothesis that
the turbulent viscosity depends on the energy at the cutof\/f.
 Knowing its relation with the energy density in Fourier space,
 they use the second order structure function and propose the {\it structure function model}:
\begin{gather}
 \ttau_s^d=C_{SF}\ov{\delta}\sqrt{\ov{F}_2(\ov{\delta}}) \, \ovs,
\label{structure}
\end{gather}
where $\ov{F}_2$ is the spatial average of the f\/iltered
structure function:
\[
 r\ \mapsto\ \ov{F}_2(r)=\iint_{\Vert \vt{z}\Vert=r} \Vert \ovu(\x)
-\ovu(\x+\vt{z}) \Vert^2 \, \textrm{d}\vt{z} \,\textrm{d} \x.
\]
\end{itemize}

The next category of models, which will be reminded, consists of
the gradient-type models.

\subsection{Gradient-type models}

To establish the gradient-type models, the subgrid stress tensor
is decomposed as follows:
\[
\ttau_s= \ovu\otimes \ovu - (\ \ov{\ovu\otimes\ovu} +
\ov{\ovu\otimes \u'}+\ov{\u'\otimes\ovu}+ \ov{\u'\otimes \u'}\ ).
\]
Next, each term between the brackets are written in Fourier space.
Then, the Fourier transform of the f\/ilter, which is assumed to
be Gaussian, is approximated by an appropriate function. Finally,
the inverse Fourier transform is computed. The models in this
category dif\/fer by the way in which the Fourier transform of the
f\/ilter is approximated.
\begin{itemize}\itemsep=0pt
\item If a second order Taylor series expansions according to the
f\/ilter width $\ov{\delta}$ is used in the approximation, one
has:
\begin{gather}
\ttau_s=-\cfrac{\ov{\delta}^2}{12}\ \nabla\ovu\ \tp\nabla\ovu.
\label{gradient}
\end{gather}

\item The gradient model is not dissipative enough and not
numerically stable~\cite{winckelmans01,iliescu02}. Thus, it is
generally combined to Smagorinsky model. This gives {\it Taylor
model}:
\begin{gather*}
\ttau_s=-\cfrac{\ov{\delta}^2}{12}\ \nabla\ovu\ \tp\nabla\ovu +
C\ovdelta^2\vert{\ovs}\vert\ovs.
\end{gather*}

\item The Taylor approximation of the Fourier transform of the
f\/ilter tends to accentuate the small frequencies rather than
attenuating them. Instead, a rational approximation can be
used~\cite{iliescu03b,berselli04}. This gives the following
expression of the model:
\begin{gather*}
\ttau_s=-\cfrac{\ov{\delta}^2}{12}\left(
\tsr{I_d}-\cfrac{\ov{\delta}^2}{24}\nabla^2 \right)^{-1}
[\nabla\ovu\ \tp\nabla\ovu] + C\ovdelta^2\vert{\ovs}\vert\ovs.
\end{gather*}
To avoid the inversion of the operator $\left(
\tsr{I_d}-\frac{\ov{\delta}^2}{24}\nabla^2 \right)$, $\ttau_s$ is
approximated by:
\begin{gather}
\ttau_s=-\cfrac{\ov{\delta}^2}{12}\ G_{\ovdelta}*[\nabla\ovu\
\tp\nabla\ovu] + C\ovdelta^2\vert{\ovs}\vert\ovs.
\label{rational2}
\end{gather}
$G_{\ovdelta}$ is the kernel of the Gaussian f\/ilter. The
convolution is done numerically. The model~(\ref{rational2}) is
called the {\it rational model}.
\end{itemize}

\subsection{Similarity-type models}

Models of this category are based on the hypothesis that the
statistic structure of the small scales are similar to the
statistic structure of the smallest resolved scales. Separation of
the resolved scales is done using a test f\/ilter (symbolized by
$\tld{\phantom{x}}$). The largest resolved scales are then
represented by $\tld{\ovu}$ and the smallest ones by
$\ovu-\tld{\ovu}$. From this hypothesis, we deduce the {\it
similarity model}:
\begin{gather}
  \ttau_s=\tld{\ovu}\otimes\tld{\ovu} - \tld{\ovu\otimes\ovu}.
\label{similarity}
\end{gather}
From this expression, many other models can be obtained by
multiplying by a coef\/f\/icient, by f\/iltering again the whole
expression or by mixing with a Smagorinsky-type model.

The last models that we will consider are Lund--Novikov-type
models.

\subsection[Lund-Novikov-type models]{Lund--Novikov-type models}

\begin{itemize}\itemsep=0pt
\item Lund and Novikov include the f\/iltered vorticity tensor
$\tsr{\ov{W}}=(\nabla\ovu-\tp\nabla\ovu)$ in the expression of the
subgrid model. Cayley--Hamilton theorem gives then the {\it
Lund--Novikov model} (see~\cite{sagaut04}):
\begin{gather}
  -\ttau_s^d=  C_1\ovdelta^2|\ovs|\ovs  + C_2\ovdelta^2(\ovs^2)^d + C_3\ovdelta^2\big(\tsr{\ov{W}}^2\big)^d  \nonumber\\
\phantom{-\ttau_s^d=}{} +
C_4\ovdelta^2(\ovs\,\tsr{\ov{W}}-\tsr{\ov{W}}\,\ovs)+
              C_5\ovdelta^2\cfrac{1}{|\ovs|}\big(\ovs^2\tsr{\ov{W}}-\ovs\,\tsr{\ov{W}}^2\big),
\label{lund}
\end{gather}
where the coef\/f\/icients $C_i$ depend on the invariants obtained
from $\ovs$ and $\tsr{\ov{W}}$. The expression of these
coef\/f\/icients are so complex that they are considered as
constants and evaluated with statistic techniques.

\item To reduce the computation cost of the previous model,
Kosovic brings a simplif\/ication and proposes the following
model:
\begin{gather}
  -\ttau_s^d=(C\ovdelta)^2 \left[2|\ovs|\ovs+C_1 \big(\ovs^2\big)^d + C_2(\ovs\,\tsr{\ov{W}}-\tsr{\ov{W}}\,\ovs)\right],
\label{kosovic}
\end{gather}
where the constants $C$, $C_1$ and $C_2$ are calculated using the
theory of homogeneous and isotropic turbulence.

\end{itemize}

The derivation of these models was done using dif\/ferent
hypothesis but did not take into consideration the symmetries of
Navier--Stokes equations which may then be destroyed. So, in the
next section, these models will be analysed by a symmetry
approach.

\section{Model analysis}

The (classical) symmetry groups of Navier--Stokes equations have
been investigated for some decades (see for example
\cite{danilov67,bytev72}). They are generated by the following
transformations:
\begin{itemize}\itemsep=0pt
\item\ The {\it time translations}:\hfill $(t,\x,\u,p) \mapsto
(t+a,\x,\u,p)$, \hspace{.1\textwidth}

\item\ the {\it pressure translations}:\hfill $(t,\x,\u,p) \mapsto
(t,\x,\u,p+\zeta(t))$, \hspace{.1\textwidth}

\item\ the {\it rotations}:\hfill $(t,\x,\u,p) \mapsto
(t,\tsr{R}\x,\tsr{R}\u,p)$, \hspace{.1\textwidth}

\item\ the {\it generalized Galilean transformations}:

\hfill $(t,\x,\u,p) \mapsto \left(t,\x+\vt{\alpha}(t),
\u+\dot{\vt{\alpha}}(t),p-\rho\, \x
\cdot\ddot{\vt{\alpha}}(t)\right)$, \hspace{.1\textwidth}

\item\ and the {\it first scaling transformations}:\hfill
$(t,\x,\u,p) \mapsto (\e^{2a}t,\e^a\x,\e^{-a}\u, \e^{-2a}\,p)$.
\hspace{.1\textwidth}

\end{itemize}

In these expressions, $a$ is a scalar, $\zeta$ (respectively
$\vt{\alpha}$) a scalar (resp.~vectorial) arbitrary function of
$t$ and $\tsr{R}$ a rotation matrix, i.e.\
 $\tsr{R} \tp \tsr{R}=\tsr{I_d}$ and $\det \tsr{R}=1$. The central dot ($\cdot$) stands for $\R^3$ scalar product.

If it is considered that $\nu$ can change during the
transformation (which is then an equivalence
transformation~\cite{ibragimov94c}), one has the {\it second
scaling transformations}:
\[
(t,\x,\u,p,\nu)\mapsto (t,e^a\x,e^a\u,e^{2a}p,e^{2a}\nu),
 \]
where $a$ is the parameter.

Navier--Stokes equations admit other known symmetries which do not
constitute a one-parameter symmetry group. They are
\begin{itemize}\itemsep=0pt

\item the {\it reflections}: \hfill $(t,\x,\u,p) \mapsto
(t,\tsr{\Lambda} \x,\tsr{\Lambda} \u,p)$,\hspace{.1\textwidth}

which are discrete symmetries, $\tsr{\Lambda}$ being a diagonal
matrix $\tsr{\Lambda}=\text{diag}(\iota_1,\iota_2,\iota_3)$ with
$\iota_i=\pm 1, \ \ i=1,2,3$,

\item\ and the {\it material indifference}: \hfill $(t,\x,\u,p)
\mapsto (t,\hat{\x},\hat{\u},\hat{p})$,\hspace{.1\textwidth}

in the limit of a 2D f\/low in a simply connected
domain~\cite{cantwell78}, with
\[
\hat{\x}=\tsr{R}(t)\ \x,\qquad \hat{\u}=\tsr{R}(t)\;
\u+\dot{\tsr{R}}(t)\; \x, \qquad
\hat{p}=p-3\omega\vphi+\cfrac{1}{2} \omega^2\Vert\x\Vert^2,
\]
where $\tsr{R}(t)$ is a 2D rotation matrix with angle $\omega t$,
$\omega$ an arbitrary real constant, $\vphi$ the usual 2D stream
function def\/ined by:
\[
\u=\mathrm{curl}(\vphi\vt{e}_3),
\]
$\boldsymbol{e}_3$ the unit vector perpendicular to the plane of
the f\/low and $\Vert\point\Vert$ the Euclidean norm.
\end{itemize}

We wish to analyse which of the models cited above is compatible
with these symmetries. The set of  solutions $(\u,p)$ of
Navier--Stokes equations (\ref{navierstokes}) is preserved by each
of the symmetries. We then require that the set of  solutions
$(\ovu,\ovp)$ of the f\/iltered equations
(\ref{navierstokesfiltrees}) is also preserved by all of these
transformations, since $(\ovu,\ovp)$ is expected to be a good
approximation of $(\u,p)$. More clearly, if a transformation
\[
T\ :\ (t,x,\u,p) \mapsto (\hat{t},\hat{x},\hat{u},\hat{p})
\]
is a symmetry of (\ref{navierstokes}), we require that the model
is such that the same transformation, applied to the f\/iltered
quantities:
\[
T : (t,x,\ovu,\ov{p}) \mapsto
(\hat{t},\hat{x},\hat{\ovu},\hat{\ov{p}}),
\]
is a symmetry of the f\/iltered equations
(\ref{navierstokesfiltrees}). When this condition holds, the model
will be said {\it invariant} under the relevant symmetry.

The f\/iltered equations (\ref{navierstokesfiltrees}) may have
other symmetries but with the above requirement, we may expect to
preserve certain properties of Navier--Stokes equations
(conservation laws, wall laws, exact solutions, spectra
properties, \dots) when approximating $(\u,p)$ by $(\ovu,\ovp)$.

We will use the hypothesis that test f\/ilters do not destroy
symmetry properties, i.e.\ $ \hat{\tld{\phi}}=\tld{\hat{\phi}} $
for any quantity $\phi$.

For the analysis, the symmetries of (\ref{navierstokes}) will be
grouped into four categories:
\begin{itemize}\itemsep=0pt
    \item[--] translations, containing time translations,
     pressure translations and the generalized Gali\-lean transformations,
    \item[--] rotations and ref\/lections,
  \item[--] scaling transformations,
  \item[--] material indif\/ference.
\end{itemize}
The aim is to search which models are invariant under the
symmetries within the considered category.

\subsection{Invariance under translations}

Since almost all existing models are autonomous in time and
pressure, the f\/iltered equations (\ref{navierstokesfiltrees})
remain unchanged when a time or pressure translation is applied.
Almost all models are then invariant under the time and the
pressure translations.

The generalized Galilean transformations, applied to the
f\/iltered variables,
 have the following form:
\[
(t,\x,\ovu,\ovp) \ \mapsto\
(\hat{t},\hat{x},\hat{\ovu},\hat{\ov{p}})=
\left(t,\x+\vt{\alpha}(t), \ovu+\dot{\vt{\alpha}}(t),\ovp-\rho\,
\x\boldsymbol{\cdot}\ddot{\vt{\alpha}}(t) \right).
\]
All models in Section 2, in which $\x$ and $\ovu$ are present only
through $\nabla\ovu$ are invariant since
\[
\hat{\nabla}\hat{\ovu}=\nabla\hat{\ovu}=\nabla\ovu,
\]
where $\hat{\nabla}=(\partial/\partial \hat{x}_1,\partial/\partial
\hat{x}_2, \partial/\partial \hat{x}_3). $

The remaining models, i.e.\ the dynamic and the similarity models
are also invariant because
\[
\widetilde{\hat{\ovu}\otimes \hat{\ovu}}
-\widetilde{\hat{\ovu}}\otimes\widetilde{\hat{\ovu}}=
\widetilde{(\ovu+\dot{\vt{\alpha}})\otimes
(\ovu+\dot{\vt{\alpha}})}
-\widetilde{(\ovu+\dot{\vt{\alpha}})}\otimes\widetilde{(\ovu+\dot{\vt{\alpha}})}=
\widetilde{\ovu\otimes \ovu}
-\widetilde{\ovu}\otimes\widetilde{\ovu}.
\]

\subsection[Invariance under rotations and reflections]{Invariance under rotations and ref\/lections}
\label{inv_rotation}

The rotations and the ref\/lections can be put together in a
transformation:
\[
(t,\x,\u,p) \mapsto (t,\tsr{\Upsilon}\x,\tsr{\Upsilon}\u,p)
\]
where $\tsr{\Upsilon}$ is a constant rotation or ref\/lection
matrix. This transformation, when applied to the f\/iltered
variables, is a symmetry of (\ref{navierstokesfiltrees}) if and
only if
\begin{gather}
  \hat{\ttau}_s=\tsr{\Upsilon}\ \ttau_s \tp\tsr{\Upsilon}.
\label{mod_rotation}
\end{gather}
Let us check if the models respect this condition.

\begin{itemize}\itemsep=0pt
\item For Smagorinsky model, we have:
\begin{gather}
  \hat{\nabla} \hat{\ovu}=[\nabla(\hat{\ovu})]\tp\tsr{\Upsilon}=
[\nabla(\tsr{\Upsilon}\ovu)]\tp\tsr{\Upsilon} =
\tsr{\Upsilon}[\nabla\ovu]\tp\tsr{\Upsilon}. \label{sma_rotation}
\end{gather}
This leads to the objectivity of $\ovs$:
\[ \hat{\ovs}=\tsr{\Upsilon}\ \ovs\ \tp\tsr{\Upsilon}.  \]
And since $\vert\hat{\ovs}\vert=\vert\ovs\vert$,
(\ref{mod_rotation}) is verif\/ied. Smagorinsky model is then
invariant.

\item For similarity model (\ref{similarity}), one has:
\[
\hat{\ovu}\otimes\hat{\ovu}=
(\tsr{\Upsilon}\ovu)\otimes(\tsr{\Upsilon}\ovu)
=\tsr{\Upsilon}(\ovu\otimes\ovu)\tp\tsr{\Upsilon}.
 \]
By means of these relations,  invariance can easily been deduced.

\item The same relations are suf\/f\/icient to prove  invariance
of the dynamic model since the trace remains invariant under a
change of orthonormal basis.

\item The structure function model (\ref{structure}) is invariant
because the function $\ov{F}_2$ is not altered under a rotation or
a ref\/lection.

\item Relations (\ref{sma_rotation}) can be used again to
 prove  invariance of each of the gradient-type models.

\item Finally, since
\[
\hat{\ov{\tsr{W}}}= \tsr{\Upsilon}\ \ov{\tsr{W}}\
\tp\tsr{\Upsilon},
\]
Lund--Novikov-type models are also invariant.

\end{itemize}

Any model of Section 2 is then invariant under the rotations and
the ref\/lections.

\subsection{Invariance under  scaling transformations}

The two scaling transformations can be gathered in a two-parameter
transformation which, when applied to the f\/iltered variables,
have the following expression:
\[
(t,\x,\ovu,\ovp,\nu)\ \mapsto\
\big(\e^{2a}t,\e^{ab}\x,\e^{b-a}\ovu,\e^{2b-2a}\ovp,\e^{2b}\nu\big)
.
\]
where $a$ and $b$ are the parameters. The f\/irst scaling
transformations corresponds to the case $b=0$ and the second ones
to the case $a=0$.

It can be checked that the f\/iltered equations
(\ref{navierstokesfiltrees}) are invariant under the two scaling
transformations if and only if
\begin{gather}
 \hat{\tsr{T}}_s=\e^{2b-2a}\tsr{T}_s.
\label{mod_scale}
\end{gather}
Since $\hat{\ovs}=\e^{-2a}\ovs$, this condition is equivalent to:
\begin{gather}
  \hat{\nu}_s=\e^{2b}\nu_s
\label{visc_scale}\end{gather} for a turbulent viscosity model.

\begin{itemize}
\item For Smagorinsky model, we have:
\[
\hat{\nu}_s=C_S\ovdelta^2\vert\hat{\ovs}\vert
=\e^{-2a}C_S\ovdelta^2\vert\ovs\vert =\e^{-2a}\nu_s.
\]
Condition (\ref{visc_scale}) is violated. The model is invariant
neither under the f\/irst nor under the second scaling
transformations. Note that the f\/ilter width $\ovdelta$ does not
vary since it is an external scale length and has no functional
dependence on the variables of the f\/low.

\item The dynamic procedure used in (\ref{dynamic}) restores the
scaling invariance. Indeed, it can be shown that:
\[
\hat{C}_d=\e^{2b+2a}C_d,
\]
that implies:
\[
\hat{\nu}_s=\hat{C}_d\ovdelta^2\vert\hat{\ovs}\vert=\e^{2b}C_d\ovdelta^2\vert\ovs\vert=
\e^{2b}\nu_s.
\]
The dynamic model is then invariant under the two scaling
transformation.

\item For the structure function model, we have:
\[
\hat{\ov{F}}_2=\e^{b-a}\ov{F}_2
\]
and then
\[
\hat{\nu}_s=\e^{b-a} \nu_{sm},
\]
that proves that the model is not invariant.

\item Since
\[
\hat{\nabla}\hat{\ovu}=\e^{2a}\nabla\ovu
\]
the gradient model (\ref{gradient}) violates (\ref{mod_scale}),
$\ttau_s$ varying in the following way:
\[
\hat{\ttau}_s=\e^{4a}\ttau_s.
\]
This also implies that none of the gradient-type models is
invariant.

\item It is straight forward to prove that the similarity model
(\ref{similarity}) verif\/ies (\ref{mod_scale}) and is invariant.

\item At last, Lund--Novikov-type models are not invariant because
they comprise a term similar to Smagorinsky model.
\end{itemize}

In fact, none of the models where the external length scale
$\ovdelta$ appears explicitly is invariant under the scaling
transformations. Note that the dynamic model, which is invariant
under these transformations, can be written in the following form:
\[
\ttau_s^d=\cfrac{\tr(\tsr{LN})}{\tr(\tsr{N}^2)}\ |\ovs|\ovs ,
\]
where
\[
\tsr{N}=\ \widetilde{|\ovs|\ovs} -
(\widetilde{\ovdelta}/\ovdelta)^2|\widetilde{\ovs}|\widetilde{\ovs}
.
\]
It is then the ratio $\hat{\ovdelta}/\ovdelta$ which is present in
the model but neither $\ovdelta$ alone nor $\hat{\ovdelta}$ alone.

In summary, the dynamic and the similarity models are the only
invariant models under the scaling transformations. Though,
scaling transformations have a particular importance because it is
with these symmetries that Oberlack~\cite{oberlack99b} derived
scaling laws and that \"Unal~\cite{unal94} proved the existence of
solutions of Navier--Stokes equations having Kolmogorov spectrum.

The last symmetry property of Navier--Stokes equations is the
material indif\/ference, in the limit of 2D f\/low, in a simply
connected domain.

\subsection[Material indifference]{Material indif\/ference}

The material indif\/ference corresponds to a time-dependent plane
rotation, with a compensation in the pressure term. We will not
write explicitly the dependence on time of the rotation
matrix~$\tsr{R}$.
\begin{itemize}\itemsep=0pt
\item The objectivity of $\ovs$ (see Section~\ref{inv_rotation})
directly leads to  invariance of Smagorinsky model.

\item For similarity model (\ref{similarity}), we have:
\begin{gather*}
 \hat{\ttau}_s = \tsr{R}\ttau_s\tp\tsr{R} +
                   \tsr{R}(\tld{\ovu\otimes\x}-\tld{\ovu}\otimes\tld{\x})\tp\tsr{R}  \\
\phantom{\hat{\ttau}_s =}{} + \dot{\tsr{R}}(\tld{\x\otimes\ovu}-
\tld{\x}\otimes\tld{\ovu})\tp \tsr{R} +
                   \dot{\tsr{R}}(\tld{\x\otimes\x} -\tld{\x}\otimes\tld{\x})\tp\dot{\tsr{R}} .
\end{gather*}
Consequently, if the test f\/ilter is such that
\begin{gather}
  (\widetilde{\ovu\otimes \x}-\widetilde{\ovu}\otimes \widetilde{\x})=0,  \qquad
  (\tld{\x\otimes\ovu}- \tld{\x}\otimes\tld{\ovu})=0, \qquad
  (\widetilde{\x\otimes \x}-\widetilde{\x}\otimes \widetilde{\x})=0,
\label{filter_material}
\end{gather}
then the similarity model is invariant under the material
indif\/ference. All f\/ilters do not have this property. For
instance, it can be shown~\cite{razafindralandy05e} that, for the
usual box f\/ilter, the left-hand sides of
equations~(\ref{filter_material}) do not vanish and are
respectively in $O(\widetilde{\ovdelta})$,
$O(\widetilde{\ovdelta})$ and~$O(\widetilde{\ovdelta}^2)$.

\item Under the same conditions (\ref{filter_material}) on the
test f\/ilter, the dynamic model is also invariant.

\item The structure function model is invariant if and only if
\begin{gather}
  \hat{\ov{F}}_2=\ov{F}_2.
\label{mat_structure}
\end{gather}
Let us calculate $\hat{\ov{F}}_2$. Let $\ovu_{\z}$ be the function
$\x\mapsto\ovu(\x+\z)$. Then
\begin{gather*}
 \hat{\ov{F}}_2 = \myint{\Vert\z\Vert=\ovdelta}{}{  \Vert (\RR\ovu+\dot{\RR}\x)
 -(\RR\u_{\z}+\dot{\RR}\x+\dot{\RR}\z) \Vert^2  }{\z} \\
\phantom{\hat{\ov{F}}_2}{} = \myint{\Vert\z\Vert=\ovdelta}{}{
\Vert\ovu-\ovu_{\z}-\tp\RR\dot{\RR}\z \Vert }{\z} .
\end{gather*}
Knowing that $\tp\RR\dot{\RR}\z=\omega \boldsymbol{e}_3\times\z$,
we get:
\[
\hat{\ov{F}}_2=\ov{F}_2+ 2\pi\omega^2\ovdelta^3 -2\omega
\myint{\Vert\z\Vert=\ovdelta}{}{  (\ovu-\ovu_{\z})\cdot
(\boldsymbol{e}_3\times\z)  }{\z}.
\]
Condition (\ref{mat_structure}) is violated. So, the structure
function model is not invariant under the material indif\/ference.
\item For the gradient model, we have:
\begin{gather}
  \hat{\nabla}\hat{\ovu}= \RR\,\nabla\ovu\,\tp\RR + \dot{\RR}\tp\RR, \qquad
  \tp\hat{\nabla}\hat{\ovu}= \RR\,\tp\nabla\ovu\,\tp\RR + \RR\,\tp\dot{\RR}.
\label{nabla_material}
\end{gather}
Let $\J$ be the matrix such that
$\dot{\RR}\tp\RR=-\omega\J=-\RR\,\tp\dot{\RR}$ or, in a component
form:
\[
\J=\begin{pmatrix} 0&1 \\ -1&0
\end{pmatrix}.
\]
Then,
\[
\hat{(\nabla\u\,\tp\nabla\u)}= \RR(\nabla\u\,\tp\nabla\u)\tp\RR
 +\omega\RR\nabla\ovu\,\tp\RR\J  -\omega\J\RR\,\tp\nabla\ovu\,\tp\RR +\omega^2\tsr{I_d}.
 \]
The commutativity between $\J$ and $\RR$ f\/inally leads to:
\[
\hat{\ttau}_s=\RR\,\ttau_s\,\tp\RR+
\omega\RR(\nabla\ovu\J-\J\,\tp\nabla\ovu)+\omega^2\tsr{I_d}.
\]
This proves that the gradient model is not invariant.

\item The other gradient-type models inherit the lack of
invariance of the gradient model.

\item It remains the Lund--Novikov-type models. We will begin with
Kosovic model (\ref{kosovic}) since it is simpler. The f\/irst two
terms of (\ref{kosovic}) are unchanged under the transformation.
For the f\/iltered vorticity tensor $\ovw$, it follows from
(\ref{nabla_material}) that:
\[
\hat{\ovw}=\RR\,\ovw\,\tp\RR  - \omega\J.
\]
Thus,
\[
\hat{\ovs}\ \hat{\ovw}-\hat{\ovw}\ \hat{\ovs} = \RR(\ovs\
\ovw-\ovw\ \ovs)\tp\RR - \omega\RR(\ovs\J-\J\ovs)\tp\RR
\]
using again the commutativity between $\J$ and $\RR$. As for them,
$\ovs$ and $\J$ are not commutative. In fact, using properties of
$\ovs$, it can be shown that $\ovs\J=-\J\ovs$. This implies that
\begin{gather}
   \hat{\ovs}\ \hat{\ovw}-\hat{\ovw}\ \hat{\ovs} = \RR(\ovs\ \ovw-\ovw\ \ovs)\tp\RR - 2\omega\RR\ovs\J\tp\RR.
\label{kosovic_material}
\end{gather}
This shows that Kosovic model is not invariant.

\item Lastly, consider Lund--Novikov model (\ref{lund}). We have:
\[
\hat{\ovw}^2=\RR\,\ovw^2\,\tp\RR -\omega\RR(\J\ovw+\ovw\J)\tp\RR
-\omega^2\tsr{I_d} .
\]
Since $\ovw$ is anti-symmetric and the f\/low is 2D, $\ovw$ is in
the form:
\[
\ovw=\begin{pmatrix} 0 & \ov{w} \\ -\ov{w} & 0 \end{pmatrix}.
\]
A direct calculation leads then to
\[
\J\ovw=\ovw\J=-\ov{w}\tsr{I_d}
\]
and
\[
\hat{\ovw}^2=\RR\,\ovw^2\,\tp\RR -(2\ov{w}-\omega)\omega\tsr{I_d}
.
\]

Let us see now how each term of the model (\ref{lund}) containing
$\ovw$ varies under the transformation.

From the last equation, we deduce the objectivity of $(\ovw^2)^d$:
\[
\hat{(\ovw^2)^d} = \RR\ (\ovw^2)^d\ \tp\RR.
 \]
For the fourth term of (\ref{lund}), we already have
(\ref{kosovic_material}). And for the last term,
\[
\hat{\ovs}^2\hat{\ovw} -\hat{\ovs}\,\hat{\ovw}^2
=\RR(\ovs^2\ovw-\ovs\,\ovw^2)\tp\RR -\omega\RR\ovs^2\,\tp\RR\J
-(2\ov{w}-\omega)\omega\RR\ovs\,\tp\RR.
\]
Putting these results together, we have:
\[
\hat{\ttau}_s^d= \RR\ttau_s^d\ \tp\RR -\omega\ovdelta^2\RR \left[
2C_4\ovs\J -C_5\cfrac{1}{\vert\ovs\vert}\left(
\ovs^2\J-(2\ov{w}-\omega)\omega\ovs \right)  \right] \tp\RR.
\]
We conclude that Lund--Novikov model is not invariant the material
indif\/ference. This ends the analysis.
\end{itemize}

Table \ref{result} summarizes the results of the above analysis.
``Y'' means that the model is invariant under all the symmetries
of the category,  ``\no'' the opposite and ``\ye'' that the model
is invariant if the conditions (\ref{filter_material}) on the test
f\/ilter is verif\/ied. It can be seen on this table that only two
models among the nine, the dynamic and the similarity models, are
invariant under the symmetry group of Navier--Stokes equations.
The scaling transformations, which are of a particular importance
(scaling laws, Kolmogorov spectrum, \dots) are violated by almost
all models.

\begin{table}\small \centering
\begin{tabular}{l@{\hspace{20pt}}c@{\hspace{20pt}}c@{\hspace{20pt}}c@{\hspace{20pt}}c}
            & Translations & Rotations,   & Scaling       & Material \\
            &              & ref\/lections & transformations& indif\/ference \\\\
Smagorinsky & Y            & Y          & \no            & Y            \\\\
Dynamic   & Y            & Y          & Y              & \ye          \\\\
Structure function  & Y    & Y          & \no            & \no          \\\\
Gradient    & Y            & Y          & \no            & \no          \\\\
Taylor      & Y            & Y          & \no            & \no          \\\\
Rational    & Y            & Y          & \no            & \no          \\\\
Similarity  & Y            & Y          & Y              & \ye          \\\\
Lund        & Y            & Y          & \no            & \no          \\\\
Kosovic     & Y            & Y          & \no            & \no          \\\\
\end{tabular}
\caption{Results of the model analysis.} Y=invariant, \no=not
invariant, \ye =invariant if (\ref{filter_material}) is
verif\/ied. \label{result}
\end{table}

The dynamic and the similarity models have an inconvenience that
they necessitate  use of a test f\/ilter. Rather constraining
conditions, (\ref{filter_material}), are then needed for these
models to preserve the material indif\/ference. In addition, the
dynamic model does not conform to the second law of thermodynamics
since it may induce a negative dissipation. Indeed, $\nu+\nu_{s}$
can take a negative value. To avoid it, an {\it a posteriori}
forcing is generally done. It consists of assigning to $\nu_s$ a
value slightly higher than $-\nu$:
\[
\nu_s=-\nu\ (1-\varepsilon),
\]
where $\varepsilon$ is a positive real number, small against 1.
Non-conformity to the second law of thermodynamics may be
detrimental for a model because, as it will be shown in
Appendix~B,
 consistence with this law leads to  stability of the model.

Considering  this lack of invariance of existing models and to
non-conformity with thermodynamical principles, we propose in the
next section a new way of deriving models which, on one hand,
possess the symmetry group of Navier--Stokes equations and, on the
other hand, are compatible with the second law of thermodynamics.

\section{Invariant and thermodynamically consistent models}

F\/irst, we will build a class of models which possess the
symmetries of Navier--Stokes equations and next ref\/ine this
class such that the models also satisfy the thermodynamics
requirement.

\subsection{Invariance under the symmetries}

Suppose that $\ovs\neq0$. Let $\ttau_s$ be an analytic function of
$\ovs$:
\begin{gather}
  \ttau_s=\mathcal{A}(\ovs).
\label{function}
\end{gather}
By this way,  invariance under the time, pressure and generalised
Galilean translations and under the ref\/lections is guaranteed.
From (\ref{function}), Cayley--Hamilton theorem and invariance
under the rotations lead to:
\begin{gather}
\ttau_s^d = A(\chi,\zeta)\; \ovs+B(\chi,\zeta)\adj^d\ovs,
\label{hamilton}
\end{gather}
where $\chi=\tr \ovs^2$ and $\zeta=\det \ovs$ are the invariants
of $\ovs$ (the third invariant, $\tr\ovs$, vanishes), $\adj$
stands for the operator def\/ined by
\[
(\adj\ovs)\ovs = (\det \ovs)\tsr{I_d},
\]
($\adj\ovs$ is simply the comatrix of $\ovs$\,) and $A$ and $B$
are arbitrary scalar functions. Contrarily to Lund--Novikov model,
these coef\/f\/icient functions will not be taken constant.

Next, a necessary and suf\/f\/icient condition for $\ttau_s$
def\/ined by (\ref{hamilton}) to be invariant under the second
scale transformations is that $\nu$ can be factorized:
\begin{gather*}
\ttau_s^d = \nu A_0(\chi,\zeta)\; \ov{S}+ \nu
B_0(\chi,\zeta)\adj^d\ov{S} .
\end{gather*}

Lastly, $\ttau_s$ is invariant under the f\/irst scaling
transformations if
\[
\hat{\ttau}_s=e^{-2a}\ttau_s.
\]
Rewritten for $A_0$ and $B_0$, this condition becomes:
\[
A_0(e^{-4a}\chi,e^{-6a}\zeta)=A_0(\chi,\zeta)  , \qquad
 B_0(e^{-4a}\chi,e^{-6a}\zeta)=e^{2a}B_0(\chi,\zeta).
 \]
After dif\/ferentiating according to $a$ and taking $a=0$, it
follows:
\[
-4\chi \partiald{A_0}{\chi}  -6\zeta \partiald{A_0}{\zeta} = 0   ,
\qquad -4\chi \partiald{B_0}{\chi}  -6\zeta \partiald{B_0}{\zeta}
= 2B_0.
\]
To satisfy these equalities, one can take
\[
A_0(\chi,\zeta)=A_1 \left(\cfrac{\zeta}{\chi^{3/2}}\right), \qquad
B_0(\chi,\zeta)=\cfrac{1}{\sqrt{\chi}}\; B_1
\left(\cfrac{\zeta}{\chi^{3/2}}\right).
\]
Finally, if $v=\cfrac{\zeta}{\chi^{3/2}}$ then
\begin{gather}
\begin{mybox}
\ttau_s^d= \nu A_1 (v)\ \ov{S} + \nu \cfrac{1}{\sqrt{\chi}}\,  B_1
(v)  \adj^d\ov{S}.
\end{mybox}
\label{a1b1}
\end{gather}

A subgrid-scale model of class (\ref{a1b1}) remains then invariant
 under the symmetry transformations of Navier--Stokes equations.

In fact several authors were interested in building invariant
models for a long time. But because they did not use Lie theory,
they did not consider some symmetries such as the sca\-ling
transformations which are particularly important. Three of the few
authors who consi\-dered all the above symmetries in the modeling of
turbulence are \"Unal~\cite{unal97} and Saveliev and
Gorokhovski~\cite{saveliev05}. The present manner to build
invariant models generalises the \"Unal's one in the sense that it
introduces $\nu$ and the invariants of $\ovs$ into the models. In
addition, \"Unal used the Reynolds averaging approach (RANS)
instead of the large-eddy simulation approach (LES) for the
turbulence modelling. Saveliev and Gorokhovski
in~\cite{saveliev05} used the LES approach but derive their model
in a dif\/ferent way than in the present article.

Let us now return to considerations which are more specif\/ic to
large eddy simulation. We know that $\ttau_{s}$ represents the
energy exchange between the resolved and the subgrid scales. Then,
it generates certain dissipation. To account for the second law of
thermodynamics, we must ensure that the total dissipation remains
positive that is not always verif\/ied by models in the
literature. In order to satisfy this condition, we ref\/ine class
(\ref{a1b1}).

\subsection{Consequences of the second law of thermodynamics}

At molecular scale, the viscous constraint is:
\[
\ttau=\partiald{\psi}{\tsr{S}}.
\]
The potential $\psi=\nu\tr \tsr{S}^2$ is convex and positive that
ensures that the molecular dissipation is positive:
\[
\Phi=\tr(\ttau \tsr{S})\geq 0.
\]
The tensor $\ttau_s$ can be considered as a subgrid constraint,
generating a dissipation
\[\Phi_s=\tr(\ttau_s\ovs).\]
To preserve compatibility with the Navier--Stokes equations, we
assume that $\ttau_s$ has the same form as~$\ttau$:
\begin{gather}
\ttau_s=\partiald{\psi_s}{\ovs}. \label{potentiel}
\end{gather}
where $\psi_s$ is a potential depending on the invariants $\chi$
and $\zeta$ of $\ovs$. This hypothesis ref\/ines
class~(\ref{a1b1}) in the following way.

Since $\tr \ovs=0$, one deduces from (\ref{potentiel}):
\[
\ttau_s^d=2\partiald{\psi_s}{\chi}\ov{S} +
\partiald{\psi_s}{\zeta}\adj^d \ov{S}.
\]
Comparing it with (\ref{a1b1}), one gets:
\[
\cfrac{1}{2}\;  \nu A_1(v)=\partiald{\psi_s}{\chi}, \qquad \nu
\cfrac{1}{\sqrt{\chi}}\, B_1(v)=\partiald{\psi_s}{\zeta}.
\]
This leads to:
\[
\partiald{}{\zeta} \left(\cfrac{1}{2}\; A_1(v)\right) =
\partiald{}{\chi} \left(\cfrac{1}{\sqrt{\chi}} B_1(v)\right).
\]
If $g$ is a primitive of $B_1$, a solution of this equation is
\begin{gather}
A_1(v)=2g(v)-3vg'(v) \qquad \text{and} \qquad B_1(v)=g'(v).
\label{g}
\end{gather}
Then, the hypothesis (\ref{potentiel}) involves  existence of a
function  $g$ such that:
\begin{gather}
\begin{mybox}
\ttau_s^d= \nu\big[2 g(v)-3vg'(v)\big] \ov{S}+
\nu\cfrac{1}{\sqrt{\chi}}\; g'(v) \adj^d\ov{S}.
\end{mybox}
\label{final}\end{gather}

Now, let $\Phi_T$ be the total dissipation. We have:
\[
\Phi_T=\tr[(\ov{\ttau}+\ttau_s)\ovs].
\]
Using (\ref{a1b1}) et (\ref{g}), one can show that
\begin{gather}
\Phi_T \geq 0  \ \Longleftrightarrow \ 1 + A_1(v)+3vB_1(v)\geq 0 \nonumber\\
\phantom{\Phi_T \geq 0  \ }{} \Longleftrightarrow \ 1 + g(v)\geq
0. \label{condition}
\end{gather}

In summary, a model belonging to class (\ref{final}) with a
continuous function $g$ verifying
\begin{gather}\begin{mybox}
1+g\geq 0,
\end{mybox}\label{inverse}
\end{gather}
is a model possessing the symmetry group of Navier--Stokes
equations conform to the second law of thermodynamics. Such a
model can  take into account the inverse energy cascade,
since~$\Phi_s$ can have negative values. In addition, by
putting~$\ovs$ in a diagonal form, it can be shown that $v$
belongs to a bounded interval $[-v^*,v^*]$ where $v^*\simeq0.136$.
Consequently, it is not necessary to satisfy (\ref{inverse}) out
of  this interval. Another important property of such a model is
its stability, in the sense that the $L^2$-norm of the f\/iltered
velocity remains bounded. In fact, all models which are consistent
with the second law of thermodynamics are stable. This will be
proved in Appendix~B.

In the next section, we show that our approach can lead to
 numerically ef\/f\/icient results. A~very simple model of
 class (\ref{final}) is then chosen and compared to the
 two most popular models which are Smagorinsky and the dynamic models (see~\cite{sagaut04,lilly92}).

\section{Numerical test}

We choose a simple linear function for $g$:
\[
g(v)=Cv.
\]
where the constant $C$ can depend on the f\/ilter width and other
parameters.

Let $d$ be the ratio:
\[
d=\cfrac{\ovdelta}{\ell},
\]
where $\ell$ is a length scale related to the size of the domain.
The introduction of this ratio is also useful to have the right
dimensions. We now take:
\[
C=(C_sd)^2,
\]
where $C_s$ is a pure constant, set to be equal to Smagorinsky
constant, i.e.\
 $C_s\simeq 0.16$. Doing so, condition (\ref{condition}) is verif\/ied and one has:
\[
\ttau_s^d= \nu (C_sd)^2 \left( -\cfrac{\det\ovs}{||\ovs||^3}\;
\ovs+ \cfrac{1}{||\ovs||}\, \adj^d\ovs \right).
\]

In the present paper, this model will be called ``invariant
model''. We use this model to simulate a f\/low within a
ventilated room (Nielsen's cavity \cite{nielsen78}) which
interests us particularly for applications in building f\/ield.
The results will then be compared to those provided by the
Smagorinsky model and the dynamic model.

\begin{figure}[t]
\begin{minipage}[b]{7.5cm}
\centering
\includegraphics[width=6.5cm]{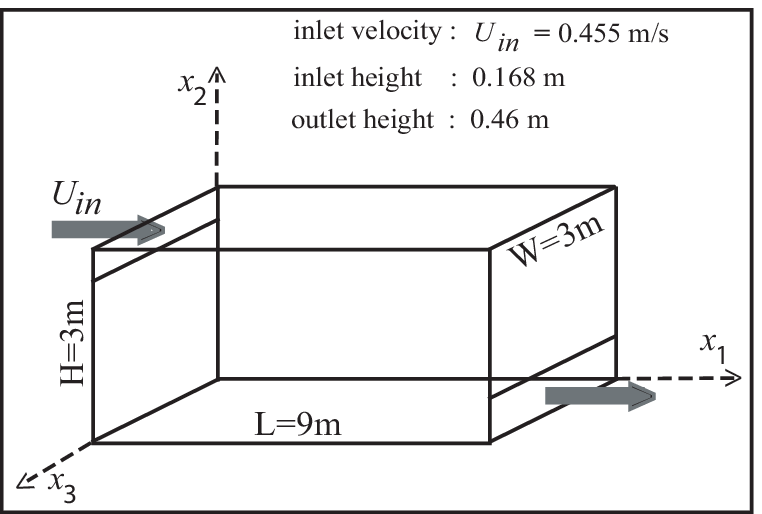}\vspace{-2mm}
\caption{Geometry of the ventilated room.}  \label{chambre}
\end{minipage}\hfill
\begin{minipage}[b]{7.5cm}
  \centering
  \includegraphics[width=6.5cm]{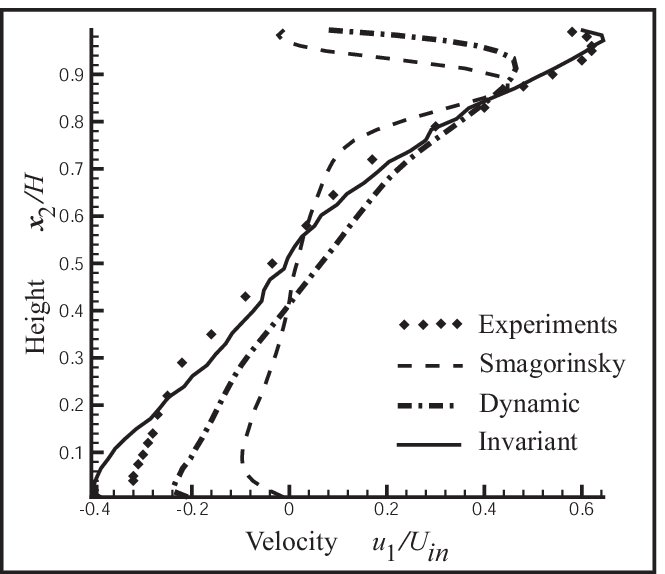}\vspace{-2mm}
  \caption{Mean velocity prof\/iles at $x/L=2/3$.}
  \label{resultat}
\end{minipage}
\end{figure}

The geometry of the room is presented on Fig.~\ref{chambre}. For
this conf\/iguration, we take $\ell=1$m.

The code used for the resolution was developed by Chen \textit{et
al} and is described in \cite{chen01}. The spatial discretization
is performed by a f\/inite dif\/ference scheme.

Fig.~\ref{resultat} compares the velocity prof\/iles given by
Smagorinsky, the dynamic and the invariant models with
experimental data at $x_1/L=2/3$ and $x_3/W=0.5$. It can be
observed on it that the invariant model gives a better result than
Smagorinsky and dynamic models, without need of a test
f\/iltering. The result is in good agreement with experiments,
except near the f\/loor. Notice that no wall model was used.

\section{Conclusion}

In this article, we presented a new class of physically compatible
subgrid turbulence models. The main ingredient used is the
symmetry group of the Navier--Stokes equations which contains a
fundamental information on the properties of the f\/low. The
second principle of thermodynamics was also introduced. From a
practical point of view,  conformity with this principle ensures
stability of the model.

A simple model of the class was tested and encouraging results was
obtained. However, the aim of this test was not to present a
complete analysis of the model but to check that the symmetry
approach can lead to good numerical results. Further studies will
be done in future works on the choice of the parameters of the
model and on  analysis of the numerical results.

The way presented here for deriving symmetry compatible models is
a general way. It can be applied to other equations
(non-isothermal f\/luid, \dots). Other parameters can also be
included. For example, dependence of the model on the viscosity
$\nu$ can be replaced by dependence on the dissipation rate.

\appendix

\section[Noether's theorem to Navier-Stokes equations]{Noether's theorem to Navier--Stokes equations}

Noether's theorem can be applied to evolution equations which can
be derived from a Lagrangian, i.e.\ evolution equations which can
be expressed in an Euler--Lagrange form:
\begin{gather}
  \partiald{L(\r)}{\r}-\Div\partiald{L(\r)}{\dot{\r}} =0,
\label{eulerlagrange}
\end{gather}
where $\r$ is the dependent variable, $\r=\r(\y)$, $\y=(y_i)_i$
the independent variable, $L$ the Lagrangian and $\Div$ the
operator:
\[
f\mapsto\Div f=\mysum{i}{}\totald{f}{y_i}.
\]
From the inf\/initesimal generators of (\ref{eulerlagrange}),
conservation laws are deduced.

Navier--Stokes equations cannot be directly written in the form
(\ref{eulerlagrange}). However, thanks to an approach of Atherton
and Homsy~\cite{atherton75}, see also \cite{ibragimov04}, which
consists in extending the Lagrangian notion, it will be shown in
this appendix that Noether's theorem can be applied to
Navier--Stokes equations.

We will say that an evolution equation
\begin{gather}
F(\y,\r)=0 \label{evolution}
\end{gather}
is derived from a ``bi-Lagrangian'' if there exists an (non
necessarily unique) application
\[
L\ :\ (\r,\s) \ \mapsto\ L(\r,\s)  \in \R
\]
such that (\ref{evolution}) is equivalent to
\[
\partiald{L(\r,\s)}{\s}-\Div\partiald{L(\r,\s)}{\dot{\s}} =0.
\]
$s$ is called the adjoint variable and the equation
\[
\partiald{L(\r,\s)}{\r}-\Div\partiald{L(\r,\s)}{\dot{\r}} =0
\]
is called the adjoint equation of (\ref{evolution}). The Noether
theorem can then be applied since the evolution equation,
associated to his adjoint, can be written in an Euler--Lagrangian
form
\[
\partiald{L(\w)}{\w}-\Div\partiald{L(\w)}{\dot{\w}} =0,
\]
where $\w=(\r,\s)$.

Navier--Stokes equations are derived from a bi-Lagrangian
\[ L((\u,p),(\v,q))=\cfrac{1}{2} \left( \partiald{\u}{t}
\scal \v-\u\scal\partiald{\v}{t} \right) + \left( q-\cfrac{1}{2}\
\u\scal \v\right) - p \div \v + \nu\tr\left(\tp\nabla
\u\scal\nabla \v\right).
 \]
where $\v=(v_i)_i$ and $q$ are the adjoint variables. The
corresponding adjoint equations are
 \begin{gather*}
 -\partiald{\v}{t}+ (\v\scal\nabla \u\tp-\u\scal\nabla \v)=\nabla q+\nu \Delta \v,\\
 \div \v=0 .
 \end{gather*}
Noether's theorem can then be applied. The inf\/initesimal
generators of the couple of equations (Navier--Stokes equations
and their adjoint equations) are:
\begin{gather*}
X_0  =  \partiald{}{t}, \\
Y_0  =  \zeta(t)\partiald{}{p},\\
X_{ij}  = x_j\partiald{}{x_i} - x_i\partiald{}{x_j} +
u_j\partiald{}{u_i} -u_i\partiald{}{u_j} +v_j\partiald{}{v_i}
-v_i\partiald{}{v_j}, \qquad i=1,2, \quad j> i,\\
X_i  =  \alpha_i(t)\partiald{}{x_i} + \alpha_i'(t)\partiald{}{u_i}
- x_i \, \alpha_i''(t)\partiald{}{p}, \qquad i=1,2,3, \\
Y_1  =  2t\partiald{}{t} + x_k\partiald{}{x_k} -u_k\partiald{}{u_k} -2p\partiald{}{p}-q\partiald{}{q} ,\\
Y'_0  =  \eta(t)\partiald{}{q},\\
X'_{ij}  = (x_ju_i-x_iu_j)\partiald{}{q} + x_j\partiald{}{v_i}
-x_i\partiald{}{v_j},
\qquad i=1,2, \quad j> i,\\
X'_i  =  \big(x_i\sigma'(t)-u_i\sigma(t)\big) \partiald{}{q}-\sigma(t)\partiald{}{v_i}, \qquad i=1,2,3,\\
Y'_1  =  v_k\partiald{}{v_k}+q\partiald{}{q},
\end{gather*}
where $\zeta$, the $\alpha_i$'s, $\eta$ and $\sigma$ are arbitrary
scalar functions.

Conservation laws for Navier--Stokes equations can be deduced from
these inf\/initesimal gene\-rators. However, that requires
non-trivial calculations and is not done in this paper.

In the last section, we will prove that a model which is
consistent with the second law of thermodynamics, i.e.\ such that
the total dissipation remains positive, is stable.

\section{Stability of thermodynamically consistent models}

After an eventual change of variables such that $\u$ vanishes
along the boundary $\Gamma$ of the domain~$\Omega$, the f\/iltered
equations can be written in the following form:
\begin{gather}
\partiald{\ovu}{t}+\div(\ovu\otimes \ovu) +\cfrac{1}{\rho}\nabla \ovp - \div (\ov{\ttau}-\ttau_s)=\F,\nonumber\\
\div \ovu=0 \label{navierstokesfiltrees2}
\end{gather}
associated to the conditions
\begin{alignat*}{3}
& \ovu=0 \qquad & & \text{sur} \ \ \Gamma, &\\
& \ovu(0,\x)=\ggamma(\x) && \text{on}\  \ \Omega, &\\
& \myint{\Omega}{}{\ovp(t,\x)}{\x}=0\qquad  && \forall \; t\in
[0,t_f].&
\end{alignat*}
$\F$ is an appropriate function of $t$ and $\x$ and $t_f$ is the
f\/inal observation time.

\medskip

\noindent {\bf Proposition.} {\it Let $(\ovu,\ovp)$ be a regular
solution of \eqref{navierstokesfiltrees2} where $\ttau_s$ is
symmetric and verif\/ies the condition:
\[
\tr[(\ov{\ttau} -\ttau_s)\ovs] \geq 0.
\]
Then:}
\[ \Vert \ovu(t,\x)\Vert _{L^2(\Omega)} \leq \Vert
\ggamma(\x)\Vert _{L^2(\Omega)} + \myint{0}{t_f}{ \Vert
\F(\tau,\x)\Vert _{L^2(\Omega)} }{\tau}.
\]

This proposition ensures a f\/inite energy when the model
conforms to the second law of thermodynamics.

\begin{proof}
Let $(\point,\point)$ denote the scalar product of $L^2(\Omega)$
and $(\ovu,\ovp)$ a regular solution of
(\ref{navierstokesfiltrees2}). From the f\/irst equation of
(\ref{navierstokesfiltrees2}) and the boundary condition, we have:
\begin{gather*}
\left(\partiald{\ovu}{t}, \ovu \right) +  b(\ovu,\ovu,\ovu) -
\cfrac{1}{\rho}\ (p,\div \ovu)+ (\ov{\ttau}-\ttau_s,\nabla \ovu) =
(\F,\ovu) ,
\end{gather*}
where $b$ is def\/ined by the trilinear form
\[
b(\u^1,\u^2,\u^3)=\big(\div(\u^1\otimes \u^2), \u^3 \big).
\]
From integrals by parts, the boundary condition and the divergence
free condition, it can be shown that $b(\ovu,\ovu,\ovu)=0$ and
$(p,\div\ovu)=0$. Since $(\ov{\ttau} -\ttau_s)$ is symmetric, it
follows that
\[
\left(\partiald{\ovu}{t}, \ovu \right) + (\ov{\ttau}
-\ttau_s,\ovs) = (\F,\ovu).
\]
Consequently
\[
\cfrac{1}{2}\ \totald{}{t}(\ovu,\ovu) + (\ov{\ttau} -\ttau_s,\ovs)
= (\F,\ovu).
\]
Now, using the main hypothesis, we have:
\[
(\ov{\ttau} -\ttau_s,\ovs)=\myint{\Omega}{}{ \tr [(\ov{\ttau}
-\ttau_s)\ovs] }{x} \geq 0
\]
and then
\[
\cfrac{1}{2}\ \totald{}{t}(\ovu,\ovu) \leq (\F,\ovu).
\]
Consequently,
\[
\Vert \ovu\Vert _{L^2(\Omega)} \totald{}{t}\Vert \ovu\Vert
_{L^2(\Omega)} \leq (\F,\ovu)
 \leq \Vert \F\Vert _{L^2(\Omega)}\Vert \ovu\Vert _{L^2(\Omega)}.
 \]
After simplifying by the $L^2$-norm of $\ovu$ and integrating over
the time, it follows:
\begin{gather*}
\Vert \ovu(t,\x)\Vert _{L^2(\Omega)} \leq \Vert \ggamma(\x)\Vert
_{L^2(\Omega)}
+ \myint{0}{t}{ \Vert \F(\tau,\x)\Vert _{L^2(\Omega)} }{\tau} \\
\phantom{\Vert \ovu(t,\x)\Vert _{L^2(\Omega)}}{} \leq \Vert
\ggamma(\x)\Vert _{L^2(\Omega)} + \myint{0}{t_f}{ \Vert
\F(\tau,\x)\Vert _{L^2(\Omega)} }{\tau}.
\end{gather*}
This ends the proof of the proposition.
\end{proof}

\LastPageEnding

\end{document}